\documentclass[conference,10pt]{IEEEtran}
\usepackage{url}
\usepackage[utf8]{inputenc}
\usepackage{color}
\usepackage{amsmath}
\usepackage{amssymb}

\usepackage[acronyms,nonumberlist,nopostdot,nomain,nogroupskip]{glossaries}
\usepackage{tablefootnote}
\usepackage{booktabs}

\usepackage{tikz}
\usepackage{pgfplots}
\pgfplotsset{compat=newest}
\pgfplotsset{plot coordinates/math parser=false}
\newlength\fheight
\newlength\fwidth
\usetikzlibrary{plotmarks,patterns,decorations.pathreplacing,backgrounds,calc,arrows,arrows.meta,spy,matrix}
\usepgfplotslibrary{patchplots,groupplots}
\usepackage{tikzscale}

\usepackage{multirow}
\usepackage{tkz-kiviat}

\usepackage{cite}

\usepackage[font=small]{subcaption}
\usepackage[font=small]{caption}

\usepackage{mathtools}

\newacronym{3gpp}{3GPP}{3rd Generation Partnership Project}
\newacronym{adc}{ADC}{Analog to Digital Converter}
\newacronym{5g}{5G}{5th generation}
\newacronym{aimd}{AIMD}{Additive Increase Multiplicative Decrease}
\newacronym{am}{AM}{Acknowledged Mode}
\newacronym{amc}{AMC}{Adaptive Modulation and Coding}
\newacronym{aqm}{AQM}{Active Queue Management}
\newacronym{awgn}{AWGN}{Additive White Gaussian Noise}
\newacronym{balia}{BALIA}{Balanced Link Adaptation}
\newacronym{bdp}{BDP}{Bandwidth-Delay Product}
\newacronym{bf}{BF}{beamforming}
\newacronym{bpsk}{BPSK}{Binary Phase-Shift Keying}
\newacronym{cc}{CC}{Congestion Control}
\newacronym{cdf}{CDF}{Cumulative Distribution Function}
\newacronym{cn}{CN}{Core Network}
\newacronym{cqi}{CQI}{Channel Quality Information}
\newacronym{cp}{CP}{Control Plane}
\newacronym{csirs}{CSI-RS}{Channel State Information - Reference Signal}
\newacronym{dc}{DC}{Dual Connectivity}
\newacronym{dce}{DCE}{Direct Code Execution}
\newacronym{dci}{DCI}{Downlink Control Information}
\newacronym{dl}{DL}{Downlink}
\newacronym{dmr}{DMR}{Deadline Miss Ratio}
\newacronym{dmrs}{DMRS}{DeModulation Reference Signal}
\newacronym{e2e}{E2E}{End-to-End}
\newacronym{ecn}{ECN}{Explicit Congestion Notification}
\newacronym{edf}{EDF}{Earliest Deadline First}
\newacronym{enb}{eNB}{evolved Node Base}
\newacronym{epc}{EPC}{Evolved Packet Core}
\newacronym{es}{ES}{Edge Server}
\newacronym{fdma}{FDMA}{Frequency Division Multiple Access}
\newacronym{fdd}{FDD}{Frequency Division Duplexing}
\newacronym[firstplural=Radio Access Technologies (RATs)]{rat}{RAT}{Radio Access Technology}
\newacronym[firstplural=Radio Access Technology (RTs)]{rt}{RT}{Radio Technology}
\newacronym{fpga}{FPGA}{Field Programmable Gate Array}
\newacronym{fs}{FS}{Fast Switching}
\newacronym{ftp}{FTP}{File Transfer Protocol}
\newacronym{gnb}{gNB}{Next Generation Node Base}
\newacronym{harq}{HARQ}{Hybrid Automatic Repeat reQuest}
\newacronym{hetnet}{HetNet}{Heterogeneous Network}
\newacronym{hh}{HH}{Hard Handover}
\newacronym{hol}{HOL}{Head-of-Line}
\newacronym{ia}{IA}{Initial Access}
\newacronym{imt}{IMT}{International Mobile Telecommunication}
\newacronym{iot}{IoT}{Internet of Things}
\newacronym{los}{LOS}{Line of Sight}
\newacronym{lte}{LTE}{Long Term Evolution}
\newacronym{m2m}{M2M}{Machine to Machine}
\newacronym{mac}{MAC}{Medium Access Control}
\newacronym{mc}{MC}{Multi-Connectivity}
\newacronym{mcs}{MCS}{Modulation and Coding Scheme}
\newacronym{mec}{MEC}{Mobile Edge Cloud}
\newacronym{mi}{MI}{Mutual Information}
\newacronym{mimo}{MIMO}{Multiple Input, Multiple Output}
\newacronym{mmib}{MMIB}{Mean Mutual Information per coded Bit}
\newacronym{mmwave}{mmWave}{millimeter wave}
\newacronym{mptcp}{MPTCP}{Multipath TCP}
\newacronym{mr}{MR}{Maximum Rate}
\newacronym{mss}{MSS}{Maximum Segment Size}
\newacronym{mtd}{MTD}{Machine-Type Device}
\newacronym{mtu}{MTU}{Maximum Transmission Unit}
\newacronym{nfv}{NFV}{Network Function Virtualization}
\newacronym{nlos}{NLOS}{Non Line of Sight}
\newacronym{nr}{NR}{New Radio}
\newacronym{ofdm}{OFDM}{Orthogonal Frequency Division Multiplexing}
\newacronym{pdcch}{PDCCH}{Physical Downlink Control Channel}
\newacronym{pdcp}{PDCP}{Packet Data Convergence Protocol}
\newacronym{pdsch}{PDSCH}{Physical Downlink Shared Channel}
\newacronym{pdu}{PDU}{Packet Data Unit}
\newacronym{pf}{PF}{Proportional Fair}
\newacronym{pgw}{PGW}{Packet Gateway}
\newacronym{phy}{PHY}{Physical}
\newacronym{pbch}{PBCH}{Physical Broadcast Channel}
\newacronym[plural=\gls{mme}s,firstplural=Mobility Management Entities (MMEs)]{mme}{MME}{Mobility Management Entity}
\newacronym{prb}{PRB}{Physical Resource Block}
\newacronym{pss}{PSS}{Primary Synchronization Signal}
\newacronym{pucch}{PUCCH}{Physical Uplink Control Channel}
\newacronym{pusch}{PUSCH}{Physical Uplink Shared Channel}
\newacronym{rach}{RACH}{Random Access Channel}
\newacronym{ran}{RAN}{Radio Access Network}
\newacronym{red}{RED}{Random Early Detection}
\newacronym{rf}{RF}{Radio Frequency}
\newacronym{rlc}{RLC}{Radio Link Control}
\newacronym{rlf}{RLF}{Radio Link Failure}
\newacronym{rrc}{RRC}{Radio Resource Control}
\newacronym{rrm}{RRM}{Radio Resource Management}
\newacronym{rr}{RR}{Round Robin}
\newacronym{rs}{RS}{Remote Server}
\newacronym{rsrp}{RSRP}{Reference Signal Received Power}
\newacronym{rss}{RSS}{Received Signal Strength}
\newacronym{rtt}{RTT}{Round Trip Time}
\newacronym{rw}{RW}{Receive Window}
\newacronym{rx}{RX}{Receiver}
\newacronym{sa}{SA}{standalone}
\newacronym{sack}{SACK}{Selective Acknowledgment}
\newacronym{sap}{SAP}{Service Access Point}
\newacronym{sc}{SC}{Single Carrier}
\newacronym{sch}{SCH}{Secondary Cell Handover}
\newacronym{scoot}{SCOOT}{Split Cycle Offset Optimization Technique}
\newacronym{sdma}{SDMA}{Spatial Division Multiple Access}
\newacronym{sinr}{SINR}{Signal to Interference plus Noise Ratio}
\newacronym{sm}{SM}{Saturation Mode}
\newacronym{snr}{SNR}{Signal to Noise Ratio}
\newacronym{son}{SON}{Self-Organizing Network}
\newacronym{ss}{SS}{Synchronization Signal}
\newacronym{srs}{SRS}{Sounding Reference Signal}
\newacronym{sss}{SSS}{Secondary Synchronization Signal}
\newacronym{tb}{TB}{Transport Block}
\newacronym{tcp}{TCP}{Transmission Control Protocol}
\newacronym{tdd}{TDD}{Time Division Duplexing}
\newacronym{tdma}{TDMA}{Time Division Multiple Access}
\newacronym{tfl}{TfL}{Transport for London}
\newacronym{tm}{TM}{Transparent Mode}
\newacronym{trp}{TRP}{Transmitter Receiver Pair}
\newacronym{tti}{TTI}{Transmission Time Interval}
\newacronym{ttt}{TTT}{Time-to-Trigger}
\newacronym{tx}{TX}{Transmitter}
\newacronym{ue}{UE}{User Equipment}
\newacronym{ul}{UL}{Uplink}
\newacronym{uml}{UML}{Unified Modeling Language}
\newacronym{um}{UM}{Unacknowledged Mode}
\newacronym{utc}{UTC}{Urban Traffic Control}
\newacronym{vm}{VM}{Virtual Machine}
\newacronym{rsrq}{RSRQ}{Reference Signal Received Quality}
\newacronym{rssi}{RSSI}{Received Signal Strength Indicator}
\newacronym{crs}{CRS}{Cell Reference Signal}
\newacronym{v2v}{V2V}{Vehicle-to-Vehicle}
\newacronym{v2i}{V2I}{Vehicle-to-Infrastructure}
\newacronym{v2x}{V2X}{Vehicle-to-Everything}
\newacronym{vn}{VN}{Vehicular Node}
\newacronym{dsrc}{DSRC}{Dedicated Short Range Communication}
\newacronym{ci}{CI}{context information}
\newacronym{voi}{VoI}{value of information}
\newacronym{gps}{GPS}{Global Positioning System}
\newacronym{qos}{QoS}{Quality of Service}
\newacronym{ml}{ML}{Machine Learning}
\newacronym{ahp}{AHP}{Analytic Hierarchy Process}
\newacronym{lidar}{LIDAR}{Light Detection and Ranging}
 \newacronym{c-its}{C-ITS}{Connected-Intelligent Transportation Systems}
 \newacronym{lan}{LAN}{Local Area Network}
\newacronym{ap}{AP}{Access Point}
\newacronym{sta}{STA}{Station}
\newacronym{cwnd}{CWND}{Congestion Window}
\newacronym{mer}{MER}{Modulation Error Ratio}
\newacronym{ttr}{TTR}{Time To Recover}
\newacronym{ber}{BER}{Bit Error Rate}
\newacronym{bler}{BLER}{Block Error Rate}
\newacronym{llr}{LLR}{Log-Likelihood Ratio}
\newacronym{per}{PER}{Packet Error Rate}
\newacronym{pfifo}{PFIFO}{Priority First In First Out}
\newacronym{dup}{DUPACK}{Duplicate Acknowledgment}
\newacronym{ack}{ACK}{Acknowledgment}
\newacronym{nack}{NACK}{Non-Acknowledgment}
\newacronym{gui}{GUI}{Graphical User Interface}
\newacronym{rto}{RTO}{Retransmission Time-Out}
\newacronym{rwnd}{RWND}{Receive Window}
\newacronym{cw}{CW}{continuous wave}
\newacronym{lsm}{LSM}{link-to-system mapping}

\tikzstyle{startstop} = [rectangle, rounded corners, minimum width=2cm, minimum height=0.5cm,text centered, draw=black]
\tikzstyle{io} = [trapezium, trapezium left angle=70, trapezium right angle=110, minimum width=3cm, minimum height=1cm, text centered, draw=black]
\tikzstyle{process} = [rectangle, minimum width=2cm, minimum height=0.5cm, text centered, draw=black, alignb=center]
\tikzstyle{decision} = [ellipse, minimum width=2cm, minimum height=1cm, text centered, draw=black]
\tikzstyle{arrow} = [thick,<->,>=stealth]
\tikzstyle{line} = [thick,>=stealth]
\tikzstyle{darrow} = [thick,<->,>=stealth,dashed]
\tikzstyle{sarrow} = [thick,->,>=stealth]
\tikzstyle{larrow} = [line width=0.1mm,dashdotted,->,>=stealth]

\makeatletter
\def\grd@save@target#1{%
  \def\grd@target{#1}}
\def\grd@save@start#1{%
  \def\grd@start{#1}}
\tikzset{
  grid with coordinates/.style={
    to path={%
      \pgfextra{%
        \edef\grd@@target{(\tikztotarget)}%
        \tikz@scan@one@point\grd@save@target\grd@@target\relax
        \edef\grd@@start{(\tikztostart)}%
        \tikz@scan@one@point\grd@save@start\grd@@start\relax
        \draw[minor help lines] (\tikztostart) grid (\tikztotarget);
        \draw[major help lines] (\tikztostart) grid (\tikztotarget);
        \grd@start
        \pgfmathsetmacro{\grd@xa}{\the\pgf@x/1cm}
        \pgfmathsetmacro{\grd@ya}{\the\pgf@y/1cm}
        \grd@target
        \pgfmathsetmacro{\grd@xb}{\the\pgf@x/1cm}
        \pgfmathsetmacro{\grd@yb}{\the\pgf@y/1cm}
        \pgfmathsetmacro{\grd@xc}{\grd@xa + \pgfkeysvalueof{/tikz/grid with coordinates/major step x}}
        \pgfmathsetmacro{\grd@yc}{\grd@ya + \pgfkeysvalueof{/tikz/grid with coordinates/major step y}}
        \foreach \x in {\grd@xa,\grd@xc,...,\grd@xb}
        \node[anchor=north] at (\x,\grd@ya) {\pgfmathprintnumber{\x}};
        \foreach \y in {\grd@ya,\grd@yc,...,\grd@yb}
        \node[anchor=east] at (\grd@xa,\y) {\pgfmathprintnumber{\y}};
      }
    }
  },
  minor help lines/.style={
    help lines,
    gray,
    line cap =round,
    xstep=\pgfkeysvalueof{/tikz/grid with coordinates/minor step x},
    ystep=\pgfkeysvalueof{/tikz/grid with coordinates/minor step y}
  },
  major help lines/.style={
    help lines,
    line cap =round,
    line width=\pgfkeysvalueof{/tikz/grid with coordinates/major line width},
    xstep=\pgfkeysvalueof{/tikz/grid with coordinates/major step x},
    ystep=\pgfkeysvalueof{/tikz/grid with coordinates/major step y}
  },
  grid with coordinates/.cd,
  minor step x/.initial=.5,
  minor step y/.initial=.2,
  major step x/.initial=1,
  major step y/.initial=1,
  major line width/.initial=1pt,
}
\makeatother

\def\code#1{\texttt{#1}}

\begin{document}


\title{QoS Provisioning in 60 GHz Communications by Physical and Transport Layer Coordination\vspace{-.3cm}}

\author{\IEEEauthorblockN{Matteo Drago$^{\circ }$, Michele Polese$^{\circ }$, Stepan Kucera$^{\dagger }$, Dmitry Kozlov$^{\dagger }$, Vitalii Kirillov$^{\dagger }$, Michele Zorzi$^{\circ }$}
\IEEEauthorblockA{
\small $^{\circ }$\small University of Padova, Padova, Italy - \texttt{\{dragomat,polesemi,zorzi\}@dei.unipd.it} \\
$^{\dagger }$\small Nokia Bell Labs, Dublin, Ireland - \texttt{\small \{stepan.kucera,dmitry.1.kozlov\}@nokia-bell-labs.com}\\\texttt{\small vitalii.1.kirillov@nokia.com}\\
}}

\makeatletter
\patchcmd{\@maketitle}
  {\addvspace{0.5\baselineskip}\egroup}
  {\addvspace{0\baselineskip}\egroup}
  {}
  {}
\makeatother

\flushbottom
\setlength{\parskip}{0ex plus0.1ex}

\maketitle

\thispagestyle{plain}
\pagestyle{plain}

\begin{abstract}
	In the last decades, technological developments in wireless communications have been coupled with an increasing demand of mobile services. From real-time applications with focus on entertainment (e.g., high quality video streaming, virtual and augmented reality), to industrial automation and security scenarios (e.g., video surveillance), the requirements are constantly pushing the limits of communication hardware and software.
	Communications at millimeter wave frequencies could provide very high throughput and low latency, thanks to the large chunks of available bandwidth, but operating at such high frequencies introduces new challenges in terms of channel reliability, which eventually impact the overall end-to-end performance.
	In this paper, we introduce a proxy that coordinates the physical and transport layers to seamlessly adapt to the variable channel conditions and avoid performance degradation (i.e., latency spikes or low throughput).
	We study the performance of the proposed solution using a simulated IEEE 802.11ad-compliant network, with the integration of input traces generated from measurements from real devices, and show that the proposed proxy-based mechanism reduces the latency by up to 50\% with respect to TCP CUBIC on a 60 GHz link.
\end{abstract}


\begin{picture}(0,0)(0,-370)
\put(0,0){
\put(0,30){\small This paper has been accepted for presentation at IEEE MASS 2019. \textcopyright[2019] IEEE.}
\put(0,20){\small Please cite it as Matteo Drago, Michele Polese, Stepan Kucera, Dmitry Kozlov, Vitalii Kirillov, Michele Zorzi, QoS Provisioning in}
\put(0,10){\small 60 GHz Communications by Physical and Transport Layer Coordination, IEEE 16th International Conference on Mobile Ad Hoc}
\put(0,0){\small and Sensor Systems (MASS), Monterey, CA, USA, 2019.}}
\end{picture}

\section{Introduction}
\label{sec:intro}
The next generations of wireless networks are being designed to address the needs and use cases of the digital society for the next decade, with an ever increasing number of connected devices and multimedia traffic that will drive the demand for wireless capacity~\cite{cisco2018cisco}. Additionally, future wireless networks will serve new verticals that so far have generally relied on wired communications, such as, for example, industrial automation, in order to provide lower deployment costs, re-configurability of the factory and flexibility in the mobility of the equipment~\cite{wang2016comparative}. This vertical introduces demanding constraints to the wireless communication stack, and a number of studies have focused on how to guarantee the combination of high reliability and low latency required on the factory floor~\cite{berardinelli2018beyond,khoshnevisan20195g}.

Recently, \glspl{mmwave} have emerged as an enabler of ultra-high data rate communications, thanks to the massive amount of bandwidth available in the spectrum between 30 and 300 GHz~\cite{rangan2017potentials}. Operations at such high frequencies are now considered in multiple standards for commercial devices, such as 3GPP NR~\cite{38300} for cellular networks and IEEE 802.11ad and 802.11ay~\cite{802.11ad-standard,802.11ay-draft} for wireless \glspl{lan}. The potential of this communication technology has also sparked interest for the aforementioned industrial automation use case. The authors of~\cite{pielli2018potential,saponara2017exploiting} consider \glspl{mmwave} as an enabling technology for future factories, thanks to the reduced interference, the small form factor of the antennas and the multi-gigabit-per-second throughput, which would allow high quality video, telemetry streaming, and low-latency sensing and actuation.

Nonetheless, communication at such high frequencies comes with a set of challenges that must be solved before the deployment in performance-critical scenarios.
The main issues are related to the intermittency of the channel, which is easily blocked and/or reflected by common materials, such as metals, brick and mortar~\cite{pi2011introduction}, as well as the human body, which can cause an attenuation that ranges from 15 dB (hand blockage)~\cite{raghavan2018statistical} to 35 dB (complete body blockage)~\cite{lu2012modeling}.
Consequently, the mobility of the communication endpoints and of the obstacles and reflectors on a factory floor may cause the channel to disappear, with sudden transitions from \gls{los} to \gls{nlos} which may happen in a time interval shorter than 100 ms~\cite{maccartney2017rapid}.
Moreover, \glspl{mmwave} are affected by a high isotropic propagation loss, which increases with the carrier frequency~\cite{pi2011introduction}.

These issues can be addressed with a combination of high density deployments, which decrease the average distance between a user and an access point and provide macro diversity~\cite{akoum2012coverage}, and directional communications, which compensate the isotropic pathloss thanks to the beamforming gain~\cite{giordani2018tutorial}. These solutions, however, have introduced a paradigm shift in the design of the wireless protocol stack, given that (i) the coverage area of a cell is now limited (thus introducing the need for smart mobility management due to the possibly high number of handovers)~\cite{polese2017jsac} and (ii) the endpoints of the communication need to track the optimal beam to be used to transmit and receive data~\cite{shokri2015mac}.

Additionally, the intermittent mmWave channel has an impact also on the higher layers of the protocol stack and, eventually, on the end-to-end performance of the network. This has been assessed by a number of simulation-based studies~\cite{zhang2016transport,zhang2019will,polese2017tcp}, which highlight how the highly variable channel and the \gls{los} to \gls{nlos} transition may affect the \gls{tcp} performance, either by reducing the throughput experienced at the application layer, thus wasting the resources available at \glspl{mmwave}, or by introducing jitter and bufferbloat (i.e., latency spikes caused by excessive buffering)~\cite{gettys2011bufferbloat}. The cross-layer interactions between the \gls{mmwave} channel and the various layers of the protocol stack thus make the end-to-end performance sub-optimal and unpredictable, impacting the \gls{qos} of the \gls{mmwave} flows. This has negative consequences not only on the user experience in mobile networks, but also, most importantly, on the reliability and end-to-end latency that this wireless technology can guarantee in an industrial automation context.

Therefore, in this paper we propose a cross-layer approach to improve the end-to-end performance of \gls{mmwave} networks, which manages to limit the latency spikes and the throughput drops and thus makes it possible to satisfy tight \gls{qos} constraints. The contributions of this paper are two-fold:
\begin{itemize}
	\item we propose a proxy design for 60 GHz wireless networks, that exploits cross-layer information to discipline the \gls{tcp} behavior, with two different control policies;
	\item we evaluate the performance of a baseline and of the proposed solution using real channel measurements at 60 GHz to drive a custom 802.11ad-based physical layer implementation in ns-3. To the best of our knowledge, this is the first study that evaluates TCP performance at 60 GHz using a mixture of simulation and experimentation, thus bridging the gap between purely testbed-based analysis (in which it is often not possible to carefully control and update the parameters of the protocol stack)~\cite{Sur:2017:WGW:3117811.3117817} and the aforementioned simulation-based evaluations~\cite{zhang2019will}.
\end{itemize}
The results show that implementing a cross-layer strategy with periodical reports could yield a reduction of up to 50\% in terms of latency, and more than double the performance in terms of throughput.

The remainder of the paper is organized as follows. In Section~\ref{sec:soa} we review the state of the art on transport layer performance in \gls{mmwave} networks. Then, in Section~\ref{sec:proxy}, we present the proxy mechanism that we introduce to enable physical and transport layer coordination. In Section~\ref{sec:setup} we describe the mixed experimental- and simulation-based evaluation setup, which is then used to obtain the results we discuss in Sec.~\ref{sec:perfeval}. Finally, we conclude the paper and suggest possible extensions in Sec.~\ref{sec:conclusions}.

\section{Transport Layer Performance at mmWaves}
\label{sec:soa}

As mentioned in Section~\ref{sec:intro}, communications at \gls{mmwave} frequencies can provide gigabit-per-second data rates at the physical layer, thanks to the large chunks of spectrum that are available in these bands~\cite{rangan2017potentials}. 

The end-to-end performance of \gls{mmwave} cellular networks has been recently in the spotlight thanks to mostly simulation studies~\cite{zhang2016transport,zhang2017tcp,polese2017tcp,kassler2017tcp,polese2017mobility,zhang2019will}, using the end-to-end mmWave module of ns-3~\cite{mezzavilla2018end}, and to some preliminary testbed-based evaluations~\cite{Sur:2017:WGW:3117811.3117817,saha2015wifi}. The studies have mainly focused on understanding the pitfalls that prevent \gls{tcp} from delivering high throughput and low latency to the application layer.
The first results were presented in~\cite{zhang2016transport,polese2017tcp}, where the authors show how TCP is too slow to react to the dynamics of the underlying \gls{mmwave} channel, because of its abstract view of the end-to-end connection. The TCP-related problems that these studies highlight are the slow ramp up of the \gls{tcp} congestion window, which leads to a sub-optimal resource utilization, the emergence of high latency spikes after \gls{los} to \gls{nlos} transitions, and the possibility of extended outages that trigger retransmission timeouts.

The first issue is linked to how the most widely used TCP congestion control algorithms (e.g., TCP CUBIC~\cite{ha2008cubic} and TCP NewReno~\cite{NewReno}) update their congestion window in the congestion avoidance phase, i.e., with a linear growth that takes too long to reach the full capacity offered by the channel~\cite{zhang2016transport}. 

The second issue is related to the appearance of bufferbloat in end-to-end connections where the \gls{ran} link is operated at \gls{mmwave} frequencies~\cite{polese2017tcp,zhang2017tcp}, with a protocol stack for the wireless link that performs buffering (as in IEEE 802.11ad~\cite{chakravarthi2016architecting} and 3GPP NR~\cite{38322}). Buffering strategies are fundamental to protect against sudden swings in the channel quality that affect the physical layer capacity, and, as highlighted in~\cite{zhang2016transport}, an undersized buffer may lead to high packet loss when the channel is in \gls{nlos} and, consequently, to severe throughput degradation. On the other hand, an oversized buffer, while protecting from these losses, thus preventing negative consequences on the throughput, at the same time makes \gls{tcp} unaware of the dynamics in the available capacity, at least until one or more packets are dropped or an \gls{aqm} procedure is triggered. Therefore, \gls{tcp} keeps sending data at a rate higher than that supported by the channel, which results in an end-to-end latency increase due to excessive buffering. As discussed in~\cite{zhang2019will}, the oversized buffer solution yields higher throughput but, at the same time, the highest average end-to-end latency for the \gls{tcp} flows. The deployment of \gls{aqm} in the buffers of the base stations represents an intermediate solution, which however does not guarantee the best performance in terms of either throughput or latency~\cite{zhang2019will}.

Finally, the third issue is caused by the possibility of extended channel outages, i.e., if there is no alternative path when the communication link is blocked. In this case, TCP may react by triggering multiple retransmission timeouts. Upon each timeout, \gls{tcp} halves the slow start threshold so that, when the link is re-established, the slow start phase (in which the congestion window grows exponentially) has a limited duration and TCP quickly transitions to congestion avoidance, thus aggravating the inefficiency associated to the slow ramp up of the congestion window.

The research community has also highlighted a number of possible solutions to allow TCP to harness the potential of \gls{mmwave} communications. In~\cite{zhang2019will}, the authors propose to increase the \gls{mss} of the \gls{tcp} flow in order to speed up the growth of the congestion window. This, however, may not be feasible in end-to-end flows that traverse Ethernet links, where the \gls{mss} should be limited to the Ethernet \gls{mtu} of 1500 bytes. Moreover, in some applications that require low latency it is not possible to aggregate enough data to create large packets. In~\cite{polese2017mobility,polese2017tcp}, multipath solutions and mobility management are used to provide macro diversity, thus decreasing the probability of a \gls{los} to \gls{nlos} transition. However, multiple backup links may not be available in all scenarios, especially when the density of the deployment is low or when the environment has few reflections. Finally, the authors of~\cite{tcp-asilomar,kim2017enhancing} propose to implement a proxy in the cellular network architecture (either in base stations or at the edge of the network) that exploits cross-layer information to steer the behavior of TCP. While~\cite{tcp-asilomar} aims at transparently forcing the \gls{tcp} congestion window to track the connection bandwidth delay product, the authors of~\cite{kim2017enhancing} also introduce an additional layer of retransmissions to protect the TCP sender from receiving duplicate acknowledgments. Finally,~\cite{azzino2017xtcp} adopts a similar approach for uplink flows, by modifying the protocol stack in the \gls{ue} which acts as TCP sender.

With respect to the the studies reviewed in this section, in this paper we assess the performance of TCP over 60 GHz links using the ns-3 simulator with real traces from an indoor environment, and propose two proxy policies that aim at enforcing \gls{qos} constraints (i.e., high throughput with low latency) over IEEE 802.11ad-based links.

\section{Physical and Transport Layer Coordination}
\label{sec:proxy}

\begin{figure}
	\centering
	\setlength\belowcaptionskip{-.3cm}
	\includegraphics[width=\columnwidth]{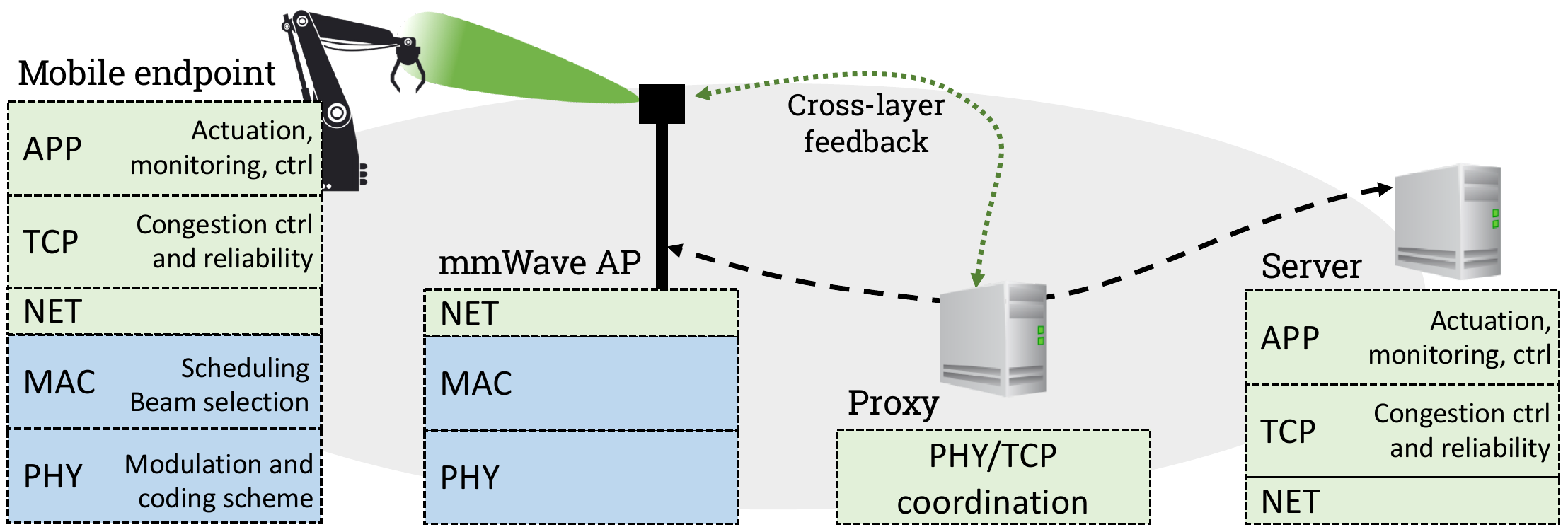}
	\caption{Factory architecture with mmWave access point and proxy.}
	\label{fig:arch}
\end{figure}

This section describes the proposed physical and transport layer coordination mechanism, along with the cross-layer policies that are implemented in the proxy. An example of the scenario of interest is shown in Figure~\ref{fig:arch}, where a \gls{mmwave} \gls{ap} at 60 GHz provides connectivity to mobile and static factory equipment, and a proxy is added to improve the performance of the end-to-end flows. The access point provides physical and \gls{mac} layer functionalities which are similar in the main protocol stacks proposed for \gls{mmwave} operations (i.e., 3GPP NR and IEEE 802.11ad/ay). In particular, the most relevant in the context of this work are \gls{amc} at the physical layer, which dictates the data rate that the wireless network offers to the higher layers, according to the experienced channel quality, and beam selection and scheduling (or medium access) at the \gls{mac} layer. 

The proxy coordinates the \gls{phy}, \gls{mac} and transport layers to efficiently exploit the available network resources. We would like to remark that the framework presented in this section is generic and can be applied to any wireless protocol stack that offers the aforementioned capabilities. In particular, the cross-layer exchange involves the following operations:
\begin{itemize}
	\item the mobile device periodically collects a matrix representing the channel quality (e.g., the received power, or \gls{snr}) over the available transmitter and receiver beam pairs $(b_{tx}, b_{rx}) \in \mathcal{B}_{tx} \times \mathcal{B}_{rx}$, where $\mathcal{B}_i$ represents the set of directions that an endpoint considers (for pilot transmission at the \gls{ap} and monitoring at the mobile device). This information is reported to the \gls{ap}. This is a standard beam management step for both 3GPP NR~\cite{giordani2018tutorial} and IEEE 802.11ad/ay~\cite{pielli2018potential};
	\item the \gls{ap} processes the report to identify the best beam to be used, as well as the \gls{mcs} that offers the best rate for a given target error rate on the link;
	\item the \gls{ap} forwards the report also to the proxy, together with additional information on which beam was chosen, the \gls{mcs}, and the scheduling policy for coordinated transmissions. This makes the proxy aware of the data rate that can be exploited on the \gls{mmwave} link, and allows the proxy to collect temporal and spatial statistics on the channel quality, that can help implement predictive policies. Moreover, the \gls{ap} may also share statistics on the buffering delays in its internal queues, if any. The reporting is done every time the \gls{bdp} estimate changes, with the aim of minimizing the communication overhead.
\end{itemize}

Moreover, contrary to~\cite{tcp-asilomar}, in this work we consider a private network deployment, in which all the equipment involved in the end-to-end flow is under the control of the private network operator (e.g., the factory owner). This is a typical scenario in the context of factory automation, as discussed in~\cite{wollschlaeger2017future}. In this configuration, the proxy can use custom control messages to interact with the TCP stack and the \gls{ap}, to perform the aforementioned procedure and, for example, gather statistics on the \gls{rtt} of the different end-to-end flows. Another deployment option in which the proxy can perform the same tasks is the traditional split configuration, in which the end-to-end connection is divided into two separate flows and the proxy optimizes the performance in the network portion towards the mobile device.

\begin{figure}[ht]
	\centering
	\setlength\belowcaptionskip{-.3cm}
	\setlength\fwidth{.9\columnwidth}
	\setlength\fheight{.6\columnwidth}
\begin{tikzpicture}
\pgfplotsset{every tick label/.append style={font=\scriptsize}}

\begin{axis}[
width=0.956\fwidth,
height=\fheight,
at={(0\fwidth,0\fheight)},
colorbar,
colorbar style={ylabel={}},
colormap = {whiteblack}{color(0cm)  = (white);color(1cm) = (black)},
point meta min=-15.4418184789447,
point meta max=16.6135134462978,
tick align=inside,
tick pos=left,
x grid style={white!69.01960784313725!black},
xlabel={RX antenna orientation [\(\displaystyle ^\circ\)]},
xlabel style={font=\footnotesize},
xmin=-0.5, xmax=12.5,
xtick style={color=black},
xtick={-1,0,1,2,3,4,5,6,7,8,9,10,11,12,13},
xticklabels={-70,-60,-50,-40,-30,-20,-10,0,10,20,30,40,50,60,70},
y grid style={white!69.01960784313725!black},
ylabel={TX antenna orientation [\(\displaystyle ^\circ\)]},
ylabel style={font=\footnotesize},
ymin=-0.5, ymax=12.5,
ytick style={color=black},
ytick={-1,0,1,2,3,4,5,6,7,8,9,10,11,12,13},
yticklabels={-70,-60,-50,-40,-30,-20,-10,0,10,20,30,40,50,60,70}
]
\addplot graphics [includegraphics cmd=\pgfimage,xmin=-0.5, xmax=12.5, ymin=12.5, ymax=-0.5] {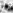};
\end{axis}

\end{tikzpicture}
	\caption{Example of \gls{snr} reporting with the matrix $\mathbf{B}_t$, with the receiver (transmitter) antenna orientation on the x (y) axis.}
	\label{fig:snapshot}
\end{figure}

Therefore, the proxy can collect the following information across the different layers of the protocol stack:

\begin{itemize}
	\item the matrix $\mathbf{B}_t$ of the \gls{snr} for each beam at time $t$, and the past matrices $\mathbf{B}_{t - i}, i \ge 1$, together with the list of beam pairs selected by the \gls{ap} and mobile device. An example of $\mathbf{B}_t$ is shown in Fig.~\ref{fig:snapshot} for the office scenario depicted in Fig.~\ref{fig:cubicle}, described in Sec.~\ref{sec:perfeval};
	\item the \gls{mcs} $M$ currently used for the communication. The \gls{mcs} usually takes discrete values (e.g., for IEEE 802.11ad it ranges from 0 to 31 \cite{802.11ad-standard}). Each value corresponds to a modulation and a coding rate, and thus can be mapped to a spectral efficiency $s$ and, eventually, to the available data rate $R = s B$, with $B$ the bandwidth allocated to the user;
	\item the minimum round-trip time $T$ of the flow that the proxy aims at optimizing.
\end{itemize}

\subsection{Proxy policies}
\label{sec:policy}
The proxy exploits the information described in the previous paragraphs to enact different flow control policies and steer the congestion window of the TCP server. In particular, we propose two different policies: (i) a reactive strategy, in which the proxy follows the dynamics of the channel and adapts accordingly; and (ii) a proactive strategy, in which the proxy tries to anticipate the evolution of the channel and possible drops in the available capacity. With both options, the proxy only modifies the value of the congestion window, while the legacy congestion awareness mechanisms (i.e., based on \gls{dup} and \gls{rto} events) are left to the TCP stack in the server~\cite{NewReno}. In this way, the overall transport protocol operations are robust against the packet loss (thanks to the retransmissions of TCP) and the erratic behavior of the channel (thanks to the cross-layer-based congestion window selection). Finally, additional policies can be developed and implemented in the proxy, and this is left as a future extension of this work.

\textbf{Reactive policy} -- In this case, the proxy receives the updates on the available rate $R$ from the \gls{ap} and, using its internal estimate of the \gls{rtt} $T$, sets the congestion window $C$ to be equal to the \gls{bdp} of the flow of interest, i.e., $C = RT$. Notice that, as discussed in~\cite{tcp-asilomar,zhang2017tcp}, by using the minimum \gls{rtt} (estimated when the system is not loaded), the \gls{bdp} is not overestimated or affected by the buffering. As outlined in the previous section, the \gls{ap} may also decide to generate a report for the proxy only when the estimated \gls{bdp} changes: this happens, for example, when the \gls{mcs} used in the communication between the \gls{ap} and the mobile device changes.

\textbf{Proactive policy} -- This scheme is similar to the reactive one, but, in addition, aims at sensing a possible imminent drop in capacity, thus giving the proxy the possibility of tuning the congestion window in advance. In this way, it is possible to reduce the excess of buffering and the increase in \gls{rtt} that happens with any reactive scheme from the moment when the channel condition changes to when the proxy receives a report related to the update. Notice that, in this paper, we consider a heuristic approach, which however can be refined with a data-driven approach based on a larger set of measurements than the one we will introduce in Section~\ref{sec:setup}.

The algorithm considers a window of $150$ ms. If during this time interval the \gls{mcs} index selected by the \gls{amc} mechanism decreases by two or more values, which is an indication of a rapid degradation of the channel quality, the algorithm enters a \textit{conservative mode}. During this phase, the TCP stack first initializes the \gls{ttr} timer, which, for the scenarios considered in Section~\ref{sec:perfeval}, is set to $0.8$ s. Then, for this entire duration, the congestion window remains set to a value equivalent to the minimum \gls{bdp}, corresponding to the lowest-rate \gls{mcs}, which protects the communication using the highest level of redundancy.
This guarantees a conservative estimate of the capacity until the \gls{ttr} expires. Once the TCP stack exits the conservative mode, the algorithm checks if the congestion window can be increased using the \gls{bdp} estimate: however, in order to avoid a sudden increase of packets that flow into the network (which would eventually lead to congested buffers), the proactive policy also ensures that the congestion window is increased gradually. If after the conservative mode the channel is still in a bad condition, the algorithm continues to rely on its conservative policies. In case an outage is experienced (i.e., temporary or permanent link breakage), the scheme interrupts the communication as long as the outage persists. It has to be further highlighted that the algorithm remains in conservative mode for the whole duration of the interval corresponding to the \gls{ttr}, even if the channel is able to recover before the timer expires. Moreover, notice that the conservative mode does not replace the acknowledgments-based recovery procedures (e.g., fast retransmission) that are implemented in the traditional versions of \gls{tcp}.
This policy is more conservative than the reactive one, thus, as we will discuss in Section~\ref{sec:perfeval}, there exists a tradeoff between the achievable throughput and the latency experienced at the application layer.

\section{Experimental and Simulation Setup}
\label{sec:setup}

This section describes the setup we used to collect the channel traces to be used in the performance evaluation.
The channel measurements (in terms of received power) have been carried out with a software-defined experimental testbed, based on \gls{fpga} boards and \gls{rf} equipment configured to work at mmWave frequencies. Each \gls{fpga} generates and modulates the digital signal representing data to transmit, which is then converted from digital to analog and up-converted to a 60 GHz carrier to be fed to a transmitting antenna.

The platform provides access to detailed information regarding the implemented hardware, such as for example:
\begin{itemize}
	\item the antenna radiation patterns and the link budget;
	\item all the physical layer parameters (e.g., modulation type, code rate, data rate) regarding each implemented \gls{mcs}.
\end{itemize}

The high level of access to the experimental platform details has been fundamental to understand the behavior of the 60 GHz channel in the scenarios of interest. For example, the information on the received power and the antenna patterns makes it possible to understand if the endpoints are communicating using the \gls{los} path or through reflections.

In a first calibration phase, we measured the \gls{mer} of actual digitally modulated transmissions \cite{hranac2006digital}. 
We performed these measurements on both \gls{los} and \gls{nlos} scenarios, where the reflections make the communication possible even without the direct path~\cite{rappaport2013millimeter}. It has to be highlighted that, in this context, the term \textit{path} refers to a particular combination of directions towards which the transmitter and receiver steer their antennas, measured by the angle between the antennas and a plane of reference \cite{akdeniz2014millimeter}. Each single \gls{mer} measurement is in fact associated to a pair of values representing the transmitter and receiver directions.

\begin{figure}
	\centering
	\begin{subfigure}[t]{.52\columnwidth}
		\centering
		\includegraphics[width=0.8\linewidth,height=0.9\linewidth]{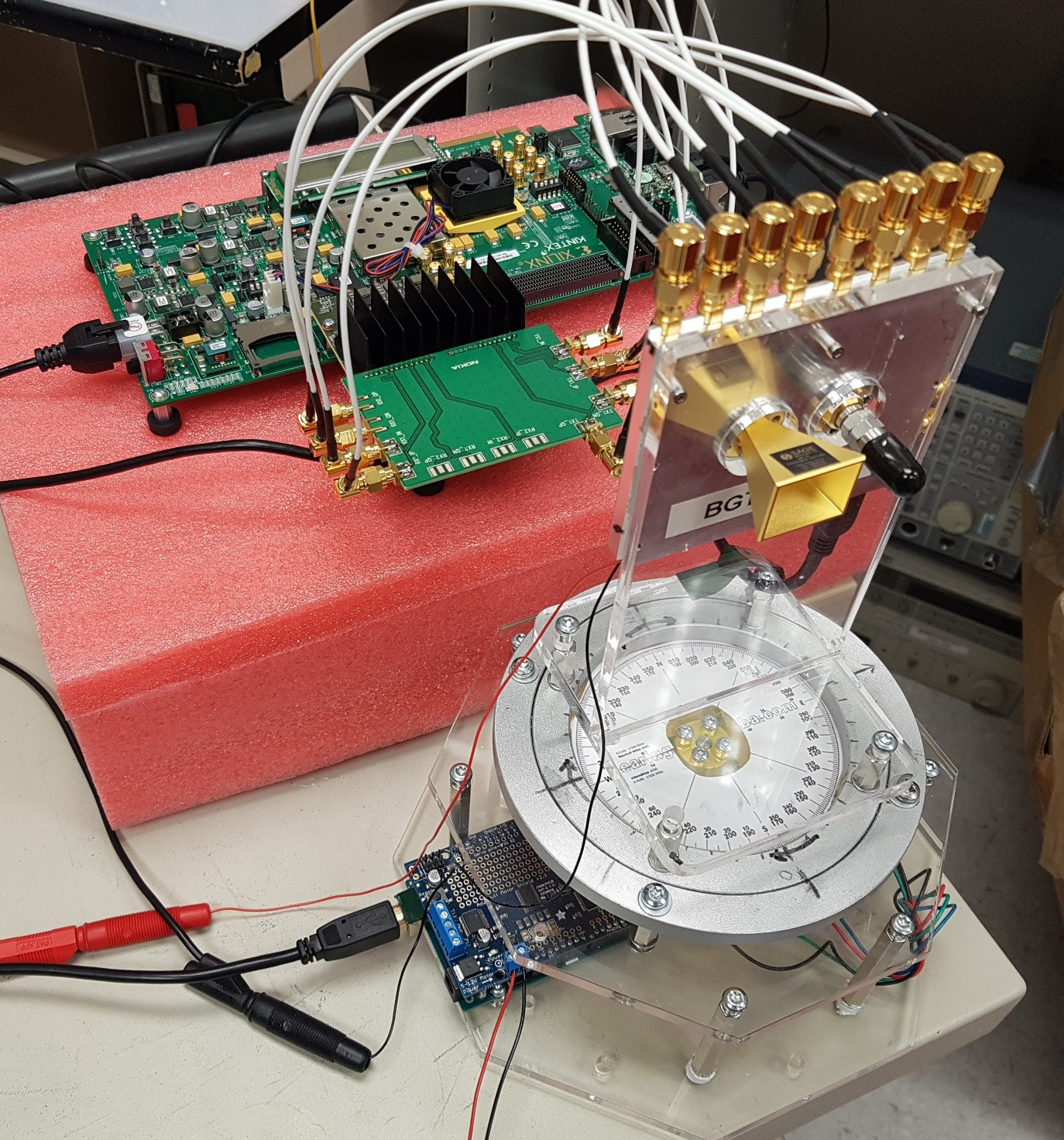}
		\caption{Transmitting device}
		\label{fig:transmitter}
	\end{subfigure}%
	\begin{subfigure}[t]{.52\columnwidth}
		\centering
		\includegraphics[width=0.8\linewidth,height=0.9\linewidth]{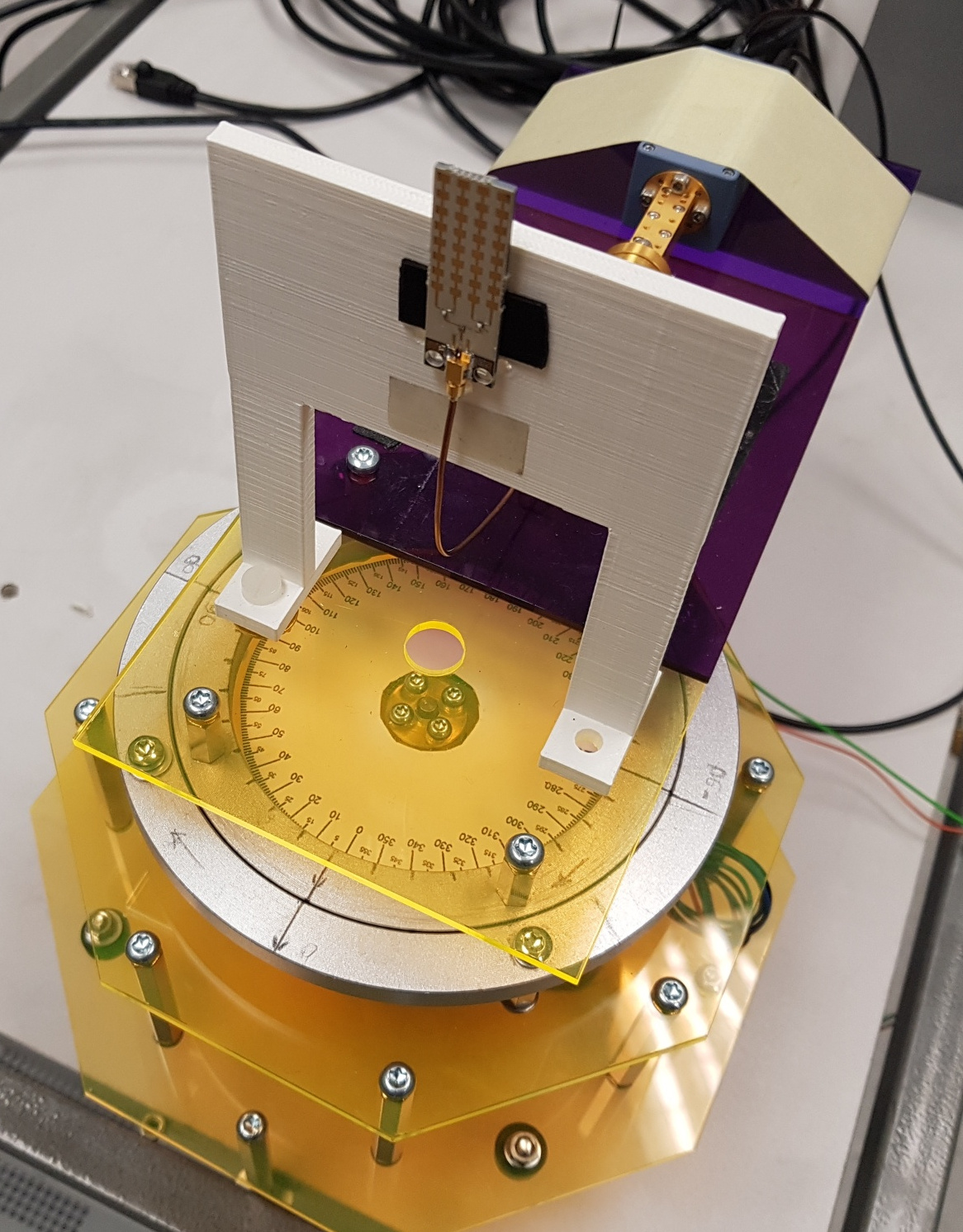}
		\caption{Receiving device}
		\label{fig:receiver}
	\end{subfigure}%
	\setlength\belowcaptionskip{-.3cm}
	\caption{Software-defined platform for the collection of the channel traces.}
	\label{fig:platform}
\end{figure}

After collecting calibration data, the second measurement phase focused on the received power from a \gls{cw}. Since the setup for this new type of measurement could give us a more precise insight on the communication link and was easier to reproduce and automate, we decided to conduct all the campaigns using the testbed designed as follows:

\begin{itemize}
	\item at the transmitter side, the \gls{fpga} is equipped with a $20$-dBi horn antenna with $14^{\circ}$ half-power beamwidth in the E-plane and $15^{\circ}$ beamwidth in the H-plane, configured to transmit a 60 GHz \gls{cw}, as shown in Figure~\ref{fig:transmitter};
	\item at the receiver side, we connected a spectrum analyzer to a $17$-dBi $32$-element patch array with $24^{\circ}$ half-power beamwidth in the E-plane and $11^{\circ}$ beamwidth in the H-plane, reported in Figure~\ref{fig:receiver}, in order to directly obtain the power of the \gls{cw}, measured in dBm.
\end{itemize}

Both the transmitting and the receiving antennas were equipped with a rotating platform automatically programmed to span an interval of directions using a pre-determined angle step of $10^{\circ}$; in addition, the acquisition procedure has been automated, in order to remove the systematic errors that could be introduced by the human operator. The output of this procedure corresponds to a matrix of values, each point representing the power measured using a specific combination of directions at the transmitter and the receiver. Notice that, for the results in Sec.~\ref{sec:perfeval}, we consider an angular space of $[-30^{\circ},30^{\circ}]$ at both endpoints.

This process was repeated for all the scenarios of interest, such as the cubicle isle of an office area and the laboratory room with different sources of reflections. In addition, starting from a specific value of \gls{cw} we devised a step-by-step algorithm to calculate the corresponding \gls{mer}, which is needed for the \gls{bler} and \gls{mcs} computation with \gls{lsm} techniques. In the following we detail each step of the algorithm, starting from the theoretical assumptions in Sec.~\ref{sec:theory}, down to the implementation of the Link-to-System mapping in Sec.~\ref{sec:lsm} and the description of the simulated network infrastructure.

\subsection{Theoretical Assumptions and Data Preprocessing}
\label{sec:theory}
Assuming to work at room temperature $T_0$ (300 K) using the same bandwidth of 500 MHz used to estimate the \gls{mer}, we modeled the noise as \gls{awgn} \cite{mueller1970capacity}. Considering that the power of the \gls{cw} is measured over a narrow bandwidth, we assume that the transceivers considered in our system can transmit and sense without distortions in the entire bandwidth. Then, applying the following formula
\begin{equation}
	P_{noise}^{CW} = 10 \log (N_0 B) + 30 \; [\mbox{dBm}],
\end{equation}
we obtain a value of $P_{noise}^{CW} = -87.01$ dBm. Substituting this value in the following equation:
\begin{equation}
\label{eq:snr}
SNR_{RX}^{CW} = P_{RX}^{CW} - P_{noise}^{CW} \; [\mbox{dB}]
\end{equation}
we can assess the \gls{snr} at the input of the receiver.


Since Eq.~\eqref{eq:snr} represents the noise-ratio up to the receiving antenna, we can express the \gls{mer} as the \gls{snr} plus an additional noise introduced by the demodulation chain. This value can then be calculated as:
\begin{equation}
\Delta = SNR_{RX}^{CW} - MER_{RX}^{FPGA} [dB],
\end{equation}
where $MER_{RX}^{FPGA}$ is the value of \gls{mer}\footnote{Notice that the \gls{mer} is the equivalent of the \gls{snr} in the digital domain, thus takes into account the imperfections of the demodulation chain.}, expressed in dB, measured at the input of the receiving \gls{fpga} board. Moreover, the measured value of \gls{mer} can be affected by several issues \cite{hranac2006digital}:
\begin{itemize}
	\item \textit{statistical variation}, which depends on the number of samples $N$; a smaller standard deviation (in general proportional to $\frac{1}{\sqrt{N}}$) means that the MER will appear more stable.
	\item \textit{nonlinear effects}, in particular on outer constellation points. It is of fundamental importance that we measure the same constellation that will be used for transmitting data; moreover, the captured sample of data must be long enough to ensure that all symbols occur with equal likelihood.
	\item \textit{\gls{mer} saturation}, which consists in the saturation of \gls{mer} at a value reflecting the implementation loss of the receiver, consisting for example in wrong symbol decoding, due to symbol detector inefficiency. 
\end{itemize}
Based on our laboratory experiments, we found that above a certain threshold the receiver front-end faces a saturation problem; this means that, even if the received power increases, the board could not yield a higher precision/lower error when demodulating the signal into a value of the constellation. For this reason we decided to set $\Delta$ to its mean value of $6.5$ dB, obtained from an extensive measurement campaign. It must be noted that the simulator module accepts $\Delta$ as a tunable parameter, in case further experiments will supply a different value.

\subsection{Link-To-System Mapping} \label{sec:lsm}
In order to map the \gls{mer} to the channel capacity and \gls{bler} for the wireless link, we designed an ad hoc platform to model the \gls{phy} layer, following the IEEE 802.11ad standard \cite{802.11ad-standard}. This module implements the procedure described in \cite{maltsev2014miweba}, which specifies for IEEE 802.11ad the approach adopted in \cite{mezzavilla2012lightweight}.

Using a suitable value of \gls{mmib}, we can evaluate the \gls{bler} as:
\begin{equation}\label{eq:bler}
BLER = \frac{1}{2} \left [ 1 - \mbox{erf}\left ( \frac{MMIB - X_1}{\sqrt{2} X_2} \right ) \right ] ,\quad X_2 \neq 0
\end{equation}
where $X_1$ and $X_2$ depend on the modulation order considered.

The complete process is the following: (i) first a value of \gls{cw} power is converted to the corresponding value of \gls{mer}; then (ii) the algorithm spans the \gls{mcs} space to find the highest value that satisfies the \gls{bler} constraint (maximum value tolerated for the wireless link), for which the actual channel \gls{bler} is evaluated as a function of \gls{mer} and the \gls{mcs} under study. The available \gls{mcs} and error rate obtained through this step-by-step procedure are assigned to the wireless channel; following previous assumptions, the channel capacity is evaluated using the spectral efficiency associated to the chosen \gls{mcs}, assuming a 500 MHz bandwidth~\cite{maltsev2014miweba}. Based on the \gls{mcs} obtained, the devices could generate channel reports, according to which congestion policy is implemented as described in Section \ref{sec:proxy}.

\subsection{Simulated Architecture}
\label{sec:simulator}
\begin{figure}
	\centering
	\includegraphics[width=0.95\columnwidth]{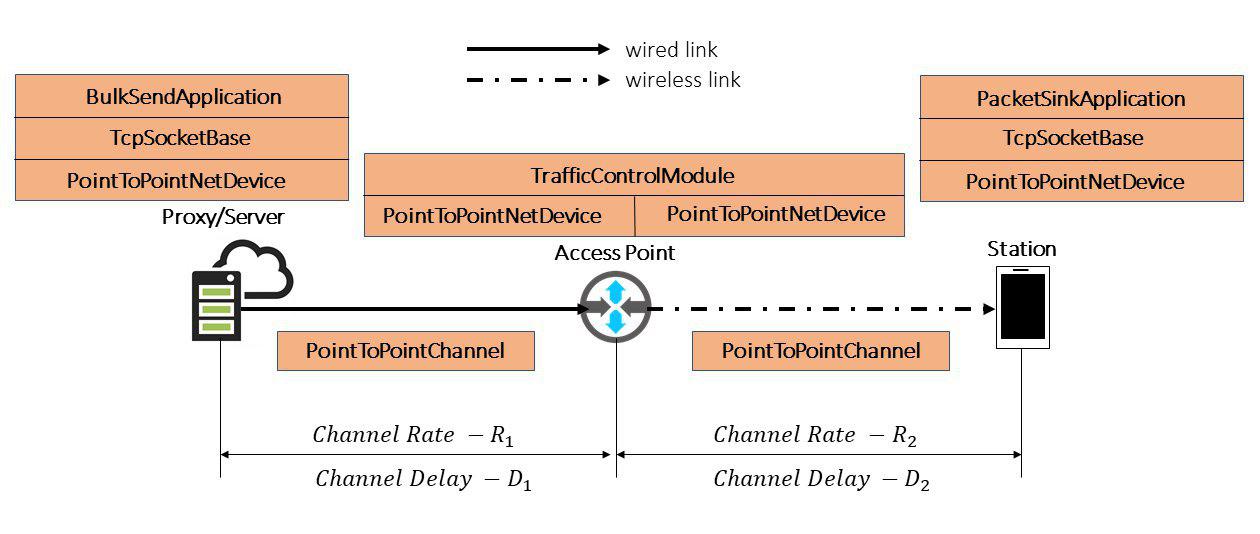}
	\caption{Network topology used in the simulation campaigns.}
	\label{fig:scenario-sim}
\end{figure}
The scenario described in Figure~\ref{fig:arch} is implemented in ns-3 using the \gls{lsm} abstraction of Section~\ref{sec:lsm} and the two-link network topology reported in Figure~\ref{fig:scenario-sim}. In this evaluation, we consider a single-user scenario. The first link includes the \gls{tcp} proxy and the \gls{ap}, which are connected using a wired link with fixed channel rate and a communication latency of $20$ ms. The second link connects the same \gls{ap}, which acts as gateway, to a receiving mobile device using a wireless link. In the following, all the modules, classes and functions that we are going to mention either are part of the ns-3 simulator's core \cite{casoni,henderson2008network} or have been designed and written during our study.

\begin{figure}
	\centering
	\begin{subfigure}[t]{.45\columnwidth}
		\centering
		\includegraphics[width=0.8\linewidth,height=0.9\linewidth]{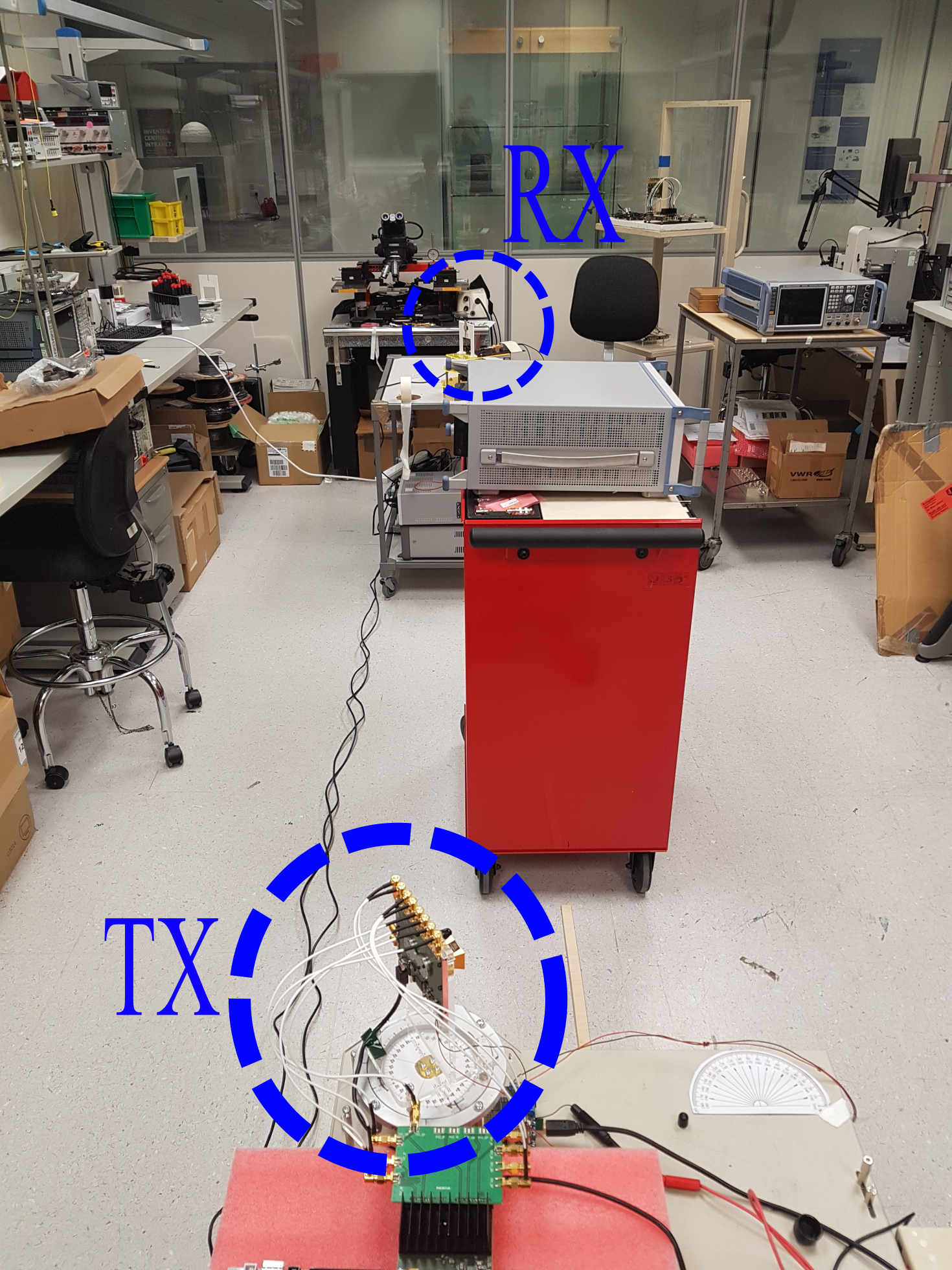}
		\caption{Scenario A.}
		\label{fig:laboratory}
	\end{subfigure}
	\begin{subfigure}[t]{.45\columnwidth}
		\centering
		\includegraphics[width=0.8\linewidth,height=0.9\linewidth]{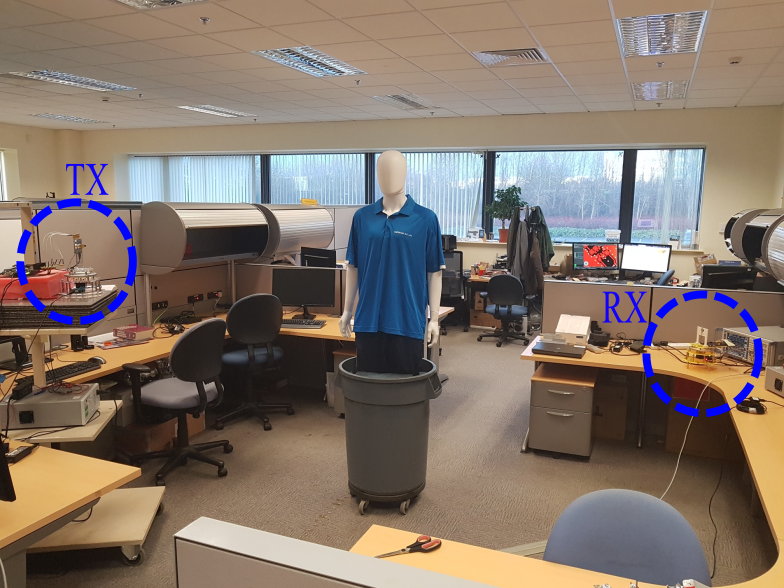}
		\caption{Scenario B.}
		\label{fig:cubicle}
	\end{subfigure}
	\caption{Framework of the measurement campaign.}
	\label{fig:scenarios}
\end{figure}

As shown in Fig.~\ref{fig:scenario-sim}, each node has installed a \code{PointToPointNetDevice} and the links between the nodes have been emulated using two distinct instances of the \code{PointToPointChannel} class. Since the \code{PointToPointChannel} does not implement any propagation loss model by default, we designed an additional module in order to associate the measured traces to the wireless link. The module consists of two main classes:
\begin{itemize}
	\item \code{MmWavePhyAbstraction}, which lists the methods and attributes used to manage the \gls{lsm} described in Section~\ref{sec:lsm};
	\item \code{MmWaveChannelTracker} which deals with the input preprocessing and provides the functions that update the wireless channel. It also manages the report infrastructure between the mobile device, the \gls{ap} and the proxy (along with the communication with the \gls{tcp} socket).
\end{itemize}
The application server continuously generates packets using \code{BulkSendApplication} with the aim of keeping the \gls{tcp} sending socket buffer filled. In this way the communication is not limited by the application behavior, and is supervised by the policies of flow and congestion control implemented in \gls{tcp}. An instance of \code{PacketSinkApplication} is installed in the receiving mobile device, which relies on the \gls{tcp} socket for the acknowledgment procedures. The \gls{ap} further implements a \code{TrafficControlModule}, provided by ns-3 and configured to use a queue implementing a \gls{pfifo} policy.

It is important to highlight that in our simulations we considered only \gls{mcs} from 1 to 12, corresponding to the \gls{sc} mode; in addition, as described in \cite{maltsev2014miweba}, in order to reach a maximum channel rate of $1.5$ Gbps, at the application layer we create packets of $8140$ Bytes. This version of the simulator, in fact, does not provide strategies of frame aggregation at the \gls{mac} layer. Also, link-layer retransmission policies have not been considered, since this study is focused on studying the performance of \gls{tcp} also in case of erratic \gls{mmwave} channel behavior and loss events. Nevertheless, this will be considered in future updates of the simulator.

\section{Performance Evaluation}
\label{sec:perfeval}

\begin{figure*}
	\centering
	\begin{subfigure}[t]{.5\textwidth}
		\setlength\fwidth{.9\textwidth}
	  	\setlength\fheight{0.5\textwidth}
		\input{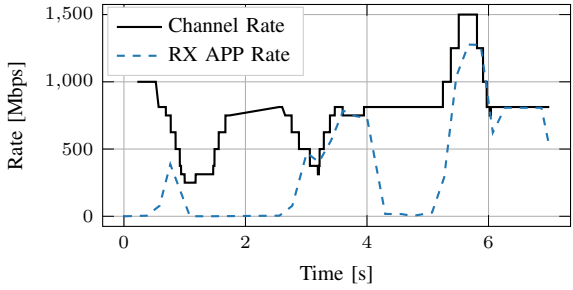}
		\caption{Channel and application rate}
		\label{fig:rate_cubic}
	\end{subfigure}%
	\hfill
	\centering
	\begin{subfigure}[t]{.5\textwidth}
		\setlength\fwidth{.9\textwidth}
	  	\setlength\fheight{0.5\textwidth}
		\input{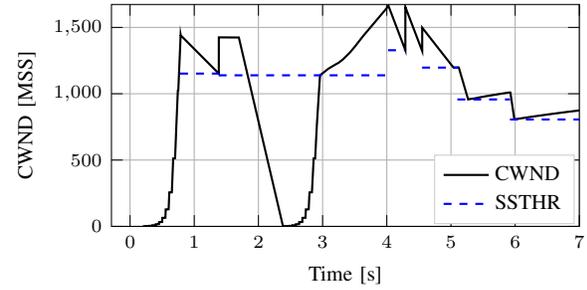}
		\caption{Congestion window and slow start threshold}
		\label{fig:cwnd_cubic}
	\end{subfigure}%
	\caption{Performance of the baseline TCP CUBIC congestion control in Scenario B.}
	\label{fig:baseline_b}
\end{figure*}

\begin{figure*}
	\centering
	\begin{subfigure}[t]{.5\textwidth}
		\setlength\fwidth{.9\textwidth}
	  	\setlength\fheight{0.5\textwidth}
		\input{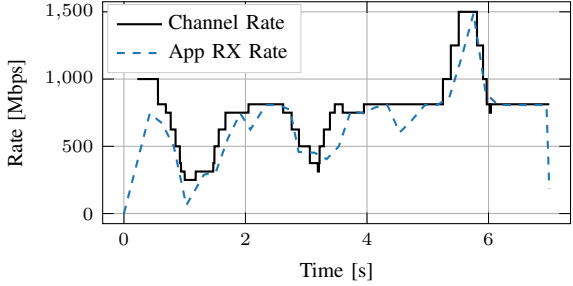}
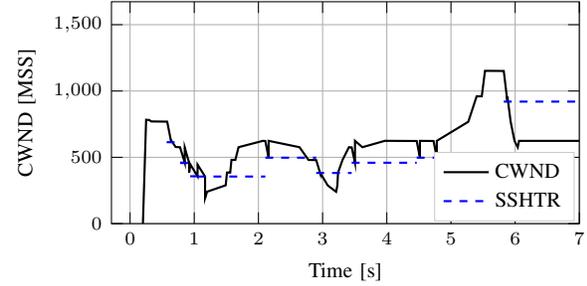
		\caption{Channel and application rate}
		\label{fig:rate_reactive}
	\end{subfigure}%
	\hfill
	\centering
	\begin{subfigure}[t]{.5\textwidth}
		\setlength\fwidth{.9\textwidth}
	  	\setlength\fheight{0.5\textwidth}
\begin{tikzpicture}
\pgfplotsset{every tick label/.append style={font=\scriptsize}}

\begin{axis}[
width=0.956\fwidth,
height=\fheight,
at={(0\fwidth,0\fheight)},
legend cell align={left},
legend style={draw=white!80.0!black},
tick align=inside,
tick pos=both,
x grid style={white!69.01960784313725!black},
xlabel={Time [s]},
xmajorgrids,
xmin=-0.2899979, xmax=7,
xtick style={color=black},
y grid style={white!69.01960784313725!black},
ylabel={CWND [MSS]},
ymajorgrids,
ymin=0, ymax=1673,
ytick style={color=black},
ylabel style={font=\footnotesize},
xlabel style={font=\footnotesize},
legend style={at={(0.99, 0.01)}, anchor=south east, font=\footnotesize}
]

\addplot [thick, black]
table {%
0.200002 1
0.250112 783.169533169533
0.300222 781.25
0.330048 770.519533169533
0.578228 769
0.641097 614
0.641178 626.047174447174
0.710078 577.889680589681
0.778672 576
0.852002 460
0.852317 573
0.920648 457
0.920779 385.259705159705
0.937036 448
1.05109 357
1.05188 445
1.16856 355
1.16882 192.629852579853
1.20002 240.787346437346
1.49008 288.94484029484
1.51009 385.259705159705
1.53018 385.165847665848
1.54004 385.085749385749
1.55008 384.590540540541
1.56006 384.546314496314
1.58006 480.682923832924
1.63008 480.558845208845
1.64007 480.186486486487
1.6901 576.22371007371
2.07009 624.242383292383
2.10728 623
2.15853 497
2.15861 624.242383292383
2.64003 576.22371007371
2.78003 480.186486486487
2.90035 479
2.96071 382
2.96084 384.14914004914
3.08001 288.111793611794
3.21003 240.093243243243
3.23002 288.111793611794
3.24009 384.14914004914
3.31009 480.186486486487
3.41006 576.22371007371
3.45306 575
3.50416 459
3.50424 624.242383292383
3.62004 576.22371007371
3.97008 624.242383292383
4.04005 624.079852579853
4.46595 623
4.51622 497
4.5163 624.079852579853
4.72864 623
4.77891 497
4.77899 624.079852579853
5.27008 768.09828009828
5.40003 960.12285012285
5.48002 960.081572481572
5.53005 1152.09791154791
5.57002 1152.00245700246
5.82246 1151
5.87931 920
5.87936 960.001965601966
5.93002 768.001597051597
5.99005 624.001228501229
6.04003 576.001228501229
6.06008 624.001228501229
7 624.001228501229
};
\addlegendentry{CWND}

\addplot [thick, blue, dashed]
table{
0.57 614
0.78 614
};\addlegendentry{SSHTR}

\addplot [thick, blue, dashed]
table{
0.78 460
0.85 460
};

\addplot [thick, blue, dashed]
table{
0.78 457
0.937036 457
};

\addplot [thick, blue, dashed]
table{
0.937036 357
1.05188 357
};

\addplot [thick, blue, dashed]
table{
1.05188 355
2.10728 355
};

\addplot [thick, blue, dashed]
table{
2.10728 497
2.90035 497
};

\addplot [thick, blue, dashed]
table{
2.9003 382
3.45306 382
};

\addplot [thick, blue, dashed]
table{
3.45306 459
4.46595 459
};

\addplot [thick, blue, dashed]
table{
4.46595 497
5.82246 497
};

\addplot [thick, blue, dashed]
table{
5.82246 920
7 920
};
\end{axis}

\end{tikzpicture}
		\caption{Congestion window and slow start threshold}
		\label{fig:cwnd_reactive}
	\end{subfigure}%
	\caption{Performance of the proxy-based reactive congestion control in Scenario B.}
	\label{fig:reactive_b}
\end{figure*}

\begin{figure}[t]
\centering
	\setlength\fwidth{.9\columnwidth}
  	\setlength\fheight{0.5\columnwidth}
	\input{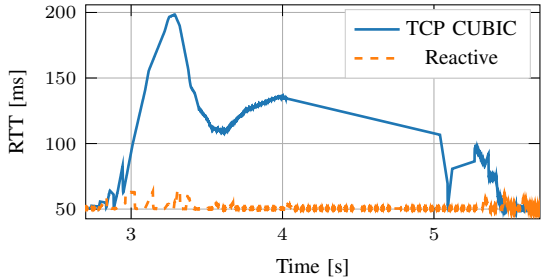}
	\caption{Latency comparison between baseline and reactive policy in Scenario B.}
	\label{fig:rtt_cmp}
\end{figure}%

The first scenario (A) we consider in this evaluation is shown in Figure~\ref{fig:laboratory}, with the antennas positioned at a height of $1$ m and at a distance of $3$ m from each other. During the measurements, a trolley is moved along a straight line in a perpendicular direction with respect to the communication path, using a $0.2$ m step. Measurements as described in Sec.~\ref{sec:setup} are taken periodically and then interpolated to create a dynamic blockage trace. In this environment, the presence of several reflectors easily provides an alternative to the main, blocked, \gls{los} path.

In the second scenario (B), shown in Fig.~\ref{fig:cubicle}, the two antennas communicate from the opposite angles of a cubicle isle: the receiver was positioned on top of a desk at a $1$ m height, while the transmitter was at a height of $1.6$ m, at a distance of $3.5$ m. In this case the \gls{los} was obstructed with a manikin of $1.9$ m of height, filled with salty water in order to experience an attenuation comparable to that of a real human being; this setup is similar to a typical indoor communication where, even if there is no blockage in the environment, there is no direct \gls{los} path available due to the different height of the devices. In this case, the blockage moved at a speed of $0.1$ m/s.

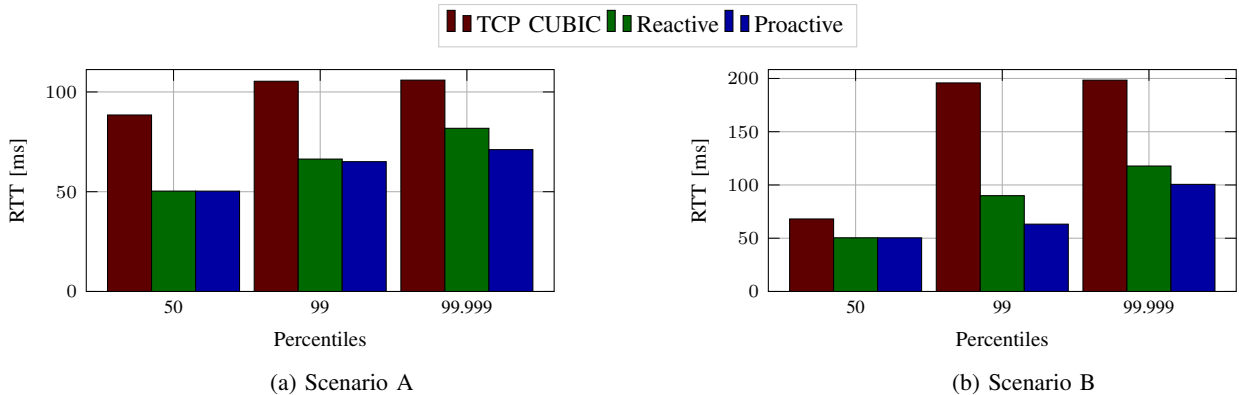
\begin{figure*}
	\centering
	\begin{subfigure}[t]{.5\textwidth}
		\setlength\abovecaptionskip{-.21cm}
		\setlength\fwidth{.9\textwidth}
	  	\setlength\fheight{0.5\textwidth}
\begin{tikzpicture}
\pgfplotsset{every tick label/.append style={font=\scriptsize}}

\begin{axis}[
width=0.956\fwidth,
height=\fheight,
at={(0\fwidth,0\fheight)},
legend cell align={left},
legend style={at={(1.2,1.1)}, anchor=south, font=\small, draw=white!80.0!black},
legend columns=3,
tick align=inside,
tick pos=both,
x grid style={white!69.01960784313725!black},
xlabel={Percentiles},
xmajorgrids,
xmin=-0.095, xmax=3.095,
xtick style={color=black},
xtick={0.5,1.5,2.5},
xticklabels={50,99,99.999},
y grid style={white!69.01960784313725!black},
ylabel={RTT [ms]},
ymajorgrids,
ymin=0, ymax=111.212859912,
ytick style={color=black},
ylabel style={font=\footnotesize},
xlabel style={font=\footnotesize},
]
\draw[draw=black,fill=red!36.86274509803922!black] (axis cs:0.05,0) rectangle (axis cs:0.35,88.4484);
\addlegendimage{ybar,ybar legend,draw=black,fill=red!36.86274509803922!black};
\addlegendentry{TCP CUBIC}

\draw[draw=black,fill=red!36.86274509803922!black] (axis cs:1.05,0) rectangle (axis cs:1.35,105.356);
\draw[draw=black,fill=red!36.86274509803922!black] (axis cs:2.05,0) rectangle (axis cs:2.35,105.91700944);
\draw[draw=black,fill=green!41.17647058823529!black] (axis cs:0.35,0) rectangle (axis cs:0.65,50.3089);
\addlegendimage{ybar,ybar legend,draw=black,fill=green!41.17647058823529!black};
\addlegendentry{Reactive}

\draw[draw=black,fill=green!41.17647058823529!black] (axis cs:1.35,0) rectangle (axis cs:1.65,66.352556);
\draw[draw=black,fill=green!41.17647058823529!black] (axis cs:2.35,0) rectangle (axis cs:2.65,81.7804118159997);
\draw[draw=black,fill=blue!63.52941176470588!black] (axis cs:0.65,0) rectangle (axis cs:0.95,50.2919);
\addlegendimage{ybar,ybar legend,draw=black,fill=blue!63.52941176470588!black};
\addlegendentry{Proactive}

\draw[draw=black,fill=blue!63.52941176470588!black] (axis cs:1.65,0) rectangle (axis cs:1.95,65.060036);
\draw[draw=black,fill=blue!63.52941176470588!black] (axis cs:2.65,0) rectangle (axis cs:2.95,71.0853548239996);
\end{axis}

\end{tikzpicture}
		\caption{Scenario A}
		\label{fig:lab_stats}
	\end{subfigure}%
	\centering
	\begin{subfigure}[t]{.5\textwidth}
		\setlength\fwidth{.9\textwidth}
	  	\setlength\fheight{0.5\textwidth}
\begin{tikzpicture}
\pgfplotsset{every tick label/.append style={font=\scriptsize}}

\begin{axis}[
width=0.956\fwidth,
height=\fheight,
at={(0\fwidth,0\fheight)},
legend cell align={left},
legend style={at={(0.5,1)}, anchor=south, draw=white!80.0!black},
legend columns=3,
tick align=inside,
tick pos=both,
x grid style={white!69.01960784313725!black},
xlabel={Percentiles},
xmajorgrids,
xmin=-0.095, xmax=3.095,
xtick style={color=black},
xtick={0.5,1.5,2.5},
xticklabels={50,99,99.999},
y grid style={white!69.01960784313725!black},
ylabel={RTT [ms]},
ymajorgrids,
ymin=0, ymax=208.3523316525,
ytick style={color=black},
ylabel style={font=\footnotesize},
xlabel style={font=\footnotesize},
]
\draw[draw=black,fill=red!36.86274509803922!black] (axis cs:0.05,0) rectangle (axis cs:0.35,68.076);
\addlegendimage{ybar,ybar legend,draw=black,fill=red!36.86274509803922!black};

\draw[draw=black,fill=red!36.86274509803922!black] (axis cs:1.05,0) rectangle (axis cs:1.35,195.80738);
\draw[draw=black,fill=red!36.86274509803922!black] (axis cs:2.05,0) rectangle (axis cs:2.35,198.43079205);
\draw[draw=black,fill=green!41.17647058823529!black] (axis cs:0.35,0) rectangle (axis cs:0.65,50.3872);
\addlegendimage{ybar,ybar legend,draw=black,fill=green!41.17647058823529!black};

\draw[draw=black,fill=green!41.17647058823529!black] (axis cs:1.35,0) rectangle (axis cs:1.65,89.9540799999999);
\draw[draw=black,fill=green!41.17647058823529!black] (axis cs:2.35,0) rectangle (axis cs:2.65,117.79428656);
\draw[draw=black,fill=blue!63.52941176470588!black] (axis cs:0.65,0) rectangle (axis cs:0.95,50.3576);
\addlegendimage{ybar,ybar legend,draw=black,fill=blue!63.52941176470588!black};

\draw[draw=black,fill=blue!63.52941176470588!black] (axis cs:1.65,0) rectangle (axis cs:1.95,63.208396);
\draw[draw=black,fill=blue!63.52941176470588!black] (axis cs:2.65,0) rectangle (axis cs:2.95,100.56948172);
\end{axis}

\end{tikzpicture}
		\caption{Scenario B}
		\label{fig:cub_stats}
	\end{subfigure}%
	\setlength\belowcaptionskip{-0.3cm}
	\caption{\gls{rtt} percentiles (50, 99, 99.999) in the two scenarios of interest.}
	\label{fig:stats}
\end{figure*}

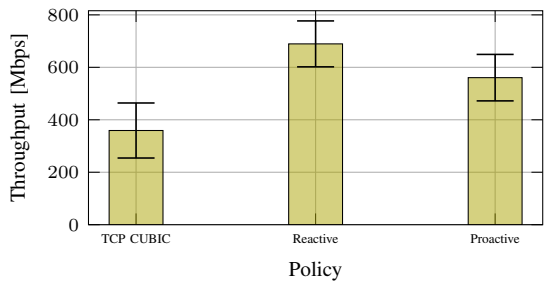
\begin{figure}[t]
\centering
	\setlength\belowcaptionskip{-0.3cm}
	\setlength\fwidth{.9\columnwidth}
  	\setlength\fheight{0.5\columnwidth}
\begin{tikzpicture}

\definecolor{color0}{rgb}{0.670588235294118,0.63921568627451,0}
\pgfplotsset{every tick label/.append style={font=\scriptsize}}

\begin{axis}[
width=0.956\fwidth,
height=\fheight,
at={(0\fwidth,0\fheight)},
tick align=inside,
tick pos=both,
x grid style={white!69.01960784313725!black},
xmajorgrids,
xmin=-0.065, xmax=2.465,
xtick style={color=black},
xtick={0.2,1.2,2.2},
xticklabels={TCP CUBIC, Reactive, Proactive},
xticklabel style = {font=\tiny},
xlabel={Policy},
y grid style={white!69.01960784313725!black},
ylabel={Throughput [Mbps]},
ymajorgrids,
ymin=0, ymax=815.780613380207,
ytick style={color=black},
ylabel style={font=\footnotesize},
xlabel style={font=\footnotesize},
]
\draw[draw=black,fill=color0,fill opacity=0.5] (axis cs:0.05,0) rectangle (axis cs:0.35,359.131281355932);
\draw[draw=black,fill=color0,fill opacity=0.5] (axis cs:1.05,0) rectangle (axis cs:1.35,689.372077777778);
\draw[draw=black,fill=color0,fill opacity=0.5] (axis cs:2.05,0) rectangle (axis cs:2.35,560.641926760563);
\path [draw=black, semithick]
(axis cs:0.2,254.215649432426)
--(axis cs:0.2,464.046913279438);

\path [draw=black, semithick]
(axis cs:1.2,601.810238050597)
--(axis cs:1.2,776.933917504959);

\path [draw=black, semithick]
(axis cs:2.2,472.051657158192)
--(axis cs:2.2,649.232196362935);

\addplot [semithick, black, mark=-, mark size=7, mark options={solid}, only marks]
table {%
0.2 254.215649432426
1.2 601.810238050597
2.2 472.051657158192
};
\addplot [semithick, black, mark=-, mark size=7, mark options={solid}, only marks]
table {%
0.2 464.046913279438
1.2 776.933917504959
2.2 649.232196362935
};
\end{axis}

\end{tikzpicture}
	\caption{Average throughput at the application layer from scenarios A and B, C.I. $95\%$.}
	\label{fig:goodput_all}
\end{figure}%

As an example, a comparison between the baseline congestion control mechanism (i.e., TCP CUBIC, the default congestion control in the Linux kernel) and the reactive policy described in Sec.~\ref{sec:policy} is shown in Figures~\ref{fig:baseline_b} and \ref{fig:reactive_b}, respectively, for scenario B.
In particular, Figure~\ref{fig:rate_cubic} highlights that the channel quality and, consequently, the available data rate at the physical layer continuously vary as the manikin moves between the two transceivers. Under these challenging conditions, the beam selection algorithm has to switch several times from one antenna sector to the other, to find the one that yields the highest received power.
As we expected, the main difference is represented by the trend of the \gls{cwnd} of the two schemes.
TCP CUBIC tries to infer the channel behavior from packet loss events but, by doing so, overestimates the \gls{cwnd} and (a) causes a massive buffer overflow, which translates into the \gls{rto} event at $\sim 2$ s in Fig.~\ref{fig:baseline_b}, and (b) generally yields a poor match between the wireless link capacity and the application layer throughput, which drops to rates smaller than 10 Mbps for extended intervals, as shown in Fig.~\ref{fig:rate_cubic}. Our reactive policy, instead, aims at exploiting the channel at its full capacity. Therefore, in the proxy-based scheme the \gls{cwnd} reflects the trend of the channel rate, as illustrated in Fig.~\ref{fig:rate_reactive}, so that the receiving application is able to achieve the rate offered by the mmWave link.

This also makes it possible to control the latency measured at the application layer, which is shown in Figure~\ref{fig:rtt_cmp} for both schemes. In particular, the results focus on an interval of time corresponding to one of the blockage events: while TCP CUBIC is affected by a high latency, due to the long queues in the buffers of the network, our scheme copes with the changes in capacity and guarantees an \gls{rtt} not higher than $70$ ms (with respect to the peak of $200$ ms experienced by TCP CUBIC).

Figure~\ref{fig:stats} reports the latency (i.e., the \gls{rtt}) for both scenarios, evaluated at different percentiles, from the median, to the worst 0.001\% of the received packets (i.e., those representing the high tail of the latency distribution of the $\sim10^6$ simulated packets). In both scenarios it is possible to observe a reduction of up to 50\% in the experienced \gls{rtt}. An additional latency reduction is obtained by using a proactive approach, which, as discussed in Sec.~\ref{sec:proxy}, makes the proxy more conservative during the intervals in which the channel is blocked, with an improvement especially on the worst percentiles. This advantage from the point of view of latency, however, comes at the cost of a lower average throughput with respect to the reactive scheme, as shown in Fig.~\ref{fig:goodput_all} (obtained by averaging the results from both scenarios). This introduces a trade-off between achievable application rate and latency, and whether to prefer one or the other largely depends on the \gls{qos} constraints dictated by the specific application.


\section{Conclusions}
\label{sec:conclusions}
In this work, we introduced a novel proxy-based approach to control and improve the performance of \gls{tcp} over \gls{mmwave} links, which, as discussed, is affected by high latency spikes and low throughput due to the erratic behavior of the channel. We proposed two congestion control policies, where the congestion window of the \gls{tcp} sender is updated according to estimates of the \acrlong{rtt} and \acrlong{bdp}. Moreover, we evaluated the performance of the proposed solution using real channel traces, collected in a number of different indoor scenarios, and fed to a custom ns-3 extension that simulates the IEEE 802.11ad physical layer.

In general, the policies that we designed outperform legacy congestion control strategies, from the point of view of both latency (compared in terms of percentiles) and average throughput (evaluated at the application layer). Moreover, we determined that the choice between our two proposed schemes must be made based on the application of interest.

These improvements come at the cost of a new specific architecture (i.e., a proxy that splits the connections) to be implemented, able to manage different transmission flows while covering all kinds of network situations. Solutions that include multiple mobile devices and access points in the same network will be considered as possible future works. Moreover, future studies will further investigate the proactive policy, using a data-driven approach.

\bibliographystyle{IEEEtran}
\bibliography{bibl.bib}

\end{document}